\documentclass[conference]{IEEEtran}
\IEEEoverridecommandlockouts

\pdfoutput=1
\usepackage{float}
\usepackage{caption}
\usepackage{subcaption}
\usepackage{amsmath}
\usepackage{comment}

\usepackage{graphicx}
\usepackage{multirow}
\usepackage{cite}
\begin{document}
%
% \title{Development of Hardware-in-Loop Framework for Satellite Communication Self-Healing Networks\\
\title{Autonomous Self-Healing UAV Swarms for Robust 6G Non-Terrestrial Networks\\
 \thanks{This research work was supported by Lockheed Martin Space.}
} 

 \author{
    \IEEEauthorblockN{Sambrama Hegde\IEEEauthorrefmark{1}, Venkata Srirama Rohit Kantheti\IEEEauthorrefmark{1}, Liang C Chu\IEEEauthorrefmark{2}, Erik Blasch\IEEEauthorrefmark{3}, and 
    Shih-Chun Lin\IEEEauthorrefmark{1}}

    \IEEEauthorblockA{\IEEEauthorrefmark{1}North Carolina State University, Raleigh, NC, USA,
     \{sambram, vkanthe, slin23\}@ncsu.edu}
     
    \IEEEauthorblockA{\IEEEauthorrefmark{2}Lockheed Martin Space, Sunnyvale, CA, USA, 
    liang.c.chu@lmco.com}
    
    \IEEEauthorblockA{\IEEEauthorrefmark{3}Air Force Research Lab, Rome, NY, USA,
    erik.blasch.1@us.af.mil}

}

\maketitle

\begin{abstract}
% Use of Non-terrestrial networks (NTNs), especially the unmanned aerial vehicles (UAVs) has been gaining interest in next-generation (next-G) wireless networks due to their potential for providing cost-effective global connectivity. This paper introduces a resilient, self-healing network designed to optimize signal quality under dynamic interference and adversarial conditions. The network leverages inter-node communication and an intelligent algorithm selection process, incorporating combining techniques like distributed-Maximal Ratio Combining (d-MRC), distributed-Linear Minimum Mean Squared Error Estimation (d-LMMSE), and Selection Combining (SC). These algorithms are selected to improve performance by adapting to changing network conditions. To evaluate the effectiveness of the proposed solution, we develop a software-defined radio (SDR)-based hardware testbed and perform detailed performance evaluations. Additionally, we present results from field tests conducted on the AERPAW testbed, which validate the proposed combining solutions in real-world scenarios. The results demonstrate that our approach significantly enhances the reliability and interference resilience of UAV networks, making them well-suited for critical communications.
Recent years have seen an increased interest in the use of Non-terrestrial networks (NTNs), especially the unmanned aerial vehicles (UAVs) to provide cost-effective global connectivity in next-generation wireless networks. We introduce a resilient, adaptive, self-healing network design (RASHND) to optimize signal quality under dynamic interference and adversarial conditions. RASHND leverages inter-node communication and an intelligent algorithm selection process, incorporating combining techniques like distributed-Maximal Ratio Combining (d-MRC), distributed-Linear Minimum Mean Squared Error Estimation (d-LMMSE), and Selection Combining (SC). These algorithms are selected to improve performance by adapting to changing network conditions. To evaluate the effectiveness of the proposed RASHND solutions, a software-defined radio (SDR)-based hardware testbed afforded initial testing and evaluations. Additionally, we present results from UAV tests conducted on the AERPAW testbed to validate our solutions in real-world scenarios. The results demonstrate that RASHND significantly enhances the reliability and interference resilience of UAV networks, making them well-suited for critical communications.
\end{abstract}


\IEEEpeerreviewmaketitle

\section{Introduction}
% With the ever increasing demand for high quality networks fueled by the need for connected devices in applications such as automation, coupled with the growth of smart cities, there has been a surge in demand for resilient networks. 
Growing demand for high quality networks, driven by connected devices, automation, and smart cities, has increased the need for resilient networks. 
% Providing adequate broadband coverage to the remote and rural areas remains a major challenge for current wireless networks.
Adequate broadband coverage in remote and rural areas remains a challenge for existing wireless networks. Increasing cellular site density can improve network resilience but entails significant deployment, operational costs, and higher energy consumption. To address these challenges, NTNs have become a key focus of next-generation (next-G) communication systems.
% Increasing the density of the cellular sites is one of the potential solutions to improve network resilience but comes with a significant deployment and operational costs and inevitably exacerbates energy consumption. 
% To address these limitations, 
% Non-terrestrial networks (NTNs) 
% NTNs have become a focal point of next generation (next-G) communication systems. 
% Moving away from the traditional infrastructure, next-G systems envision to have heterogeneous architectures where terrestrial networks are complemented by UAVs, high altitude platforms (HAPs), and satellite systems. 
Next-G systems shift from traditional infrastructure to heterogeneous architectures, integrating terrestrial networks with NTNs, and satellites.
These systems aim to offer on-demand and cost-effective coverage solutions across regions,  thereby enabling seamless, ubiquitous, and high-capacity connectivity \cite{6g_challenge, beyond_5g, giordani, ntn_azari}.

% Among the NTNs, UAVs come with the advantage of high mobility and flexible deployment, and can provide connectivity to devices without infrastructure coverage. UAVs can adapt to changing demands by acting as relays, flying base stations, or mobile edge computing nodes that enhance both network coverage and computational power. Unlike fixed networks, UAVs can access hard-to-reach areas, making them ideal for connecting rural and isolated communities 
UAVs among NTNs offer high mobility and flexible deployment, providing connectivity to areas without infrastructure. They adapt as relays, flying base stations, or mobile edge computing nodes to boost coverage and computation, reaching hard-to-reach rural and isolated areas unlike fixed networks \cite{uav}. Moreover, UAVs can operate collaboratively in swarms to form Single-Input Multiple-Output (SIMO) or Multiple-Input Multiple-Output (MIMO) systems, enabling spatial diversity and thus, improving signal quality. By
leveraging diversity in swarm configurations, the impact of
jammers can be minimized, improving the overall reliability
of the network \cite{milcom}.

Self-healing is the ability of the system to detect and recover from abnormalities with minimal or no human intervention \cite{sh1}. Self-healing network (SHNs) play an important role in complex environments like the ones involving UAVs, where having manual intervention is not feasible. SHNs help network function smoothly by handling issues including interference and node failures. A SHN intelligently switches between various combining techniques based on the condition. This ensures that the system is reliable and has high performance even in adverse conditions. 
%We intend to build a self-healing network that intelligently switches between various combining techniques based on the condition. This ensures that the system is reliable and has high performance even in adverse conditions. 

Inter-device links (IDL) are used to ensure reliable communication between the UAVs \cite{d2d}. IDLs also enable key applications like swarm flight and collision avoidance, while improving spectral/energy efficiency, coverage, and reducing backhaul dependency
% IDLs are also crucial in enabling various applications including swarm flight and collision avoidance, improving spectral and energy efficiency along with extended coverage and reduced dependency on backhaul
\cite{d2d2}. IDLs allow UAVs to exchange data by establishing direct links and thus maintain reliable connections even in challenging environments. We develop the RASHND system with device-to-device (D2D) communication functionalities to mimic the characteristics of IDLs.

% The proposed hardware end-to-end system is implemented using the Universal Software Radio Peripheral (USRP) B210's along with Raspberry Pi's (RPi) to enable self-healing on bitstream and image data. The d-MRC and  d-LMMSE algorithms used in self-healing along with the signal processing techniques used to process and extract the data have already been introduced in our previous work \cite{milcom}. The wireless transmission of the extracted data among the RPi's happens seamlessly through an internal ad-hoc network. AERPAW testbed is used to validate our results on UAV based testbed \cite{aerpaw}.

The proposed hardware end-to-end system is implemented using the Universal Software Radio Peripheral (USRP) B210s along with Raspberry Pi's (RPi) to enable self-healing on bitstream and image data. 
The d-MRC and  d-LMMSE algorithms used in self-healing along with the signal processing techniques used to process and extract the data have already been introduced in our previous work \cite{milcom}. The wireless transmission of the extracted data among the RPis happens seamlessly through an internal ad-hoc network. 
AERPAW testbed is used to validate our results on UAV based testbed \cite{aerpaw}.

The remainder of the paper is divided as follows. Section II discusses the proposed RASHND system architecture and the indoor and outdoor hardware testbeds. Section III introduces the algorithms and the state machine used for our SHN. Section IV talks about inter-device links. We present the results in section V and finally conclude the paper in section VI.

\section{System Architecture}
% This section presents the RASHND architecture along with the indoor and outdoor hardware testbed used to validate our algorithms.
This section presents the RASHND architecture, along with the hardware testbeds used to validate our algorithms.

\subsection{Proposed Architectural Model}

\begin{figure}
\centering
    %\includegraphics[width=0.8\columnwidth]{diagram1.png}
    % \includesvg[inkscapelatex=false,width=0.9\columnwidth]{images/proposed_uav.svg}
    \includegraphics[width=0.9\columnwidth]{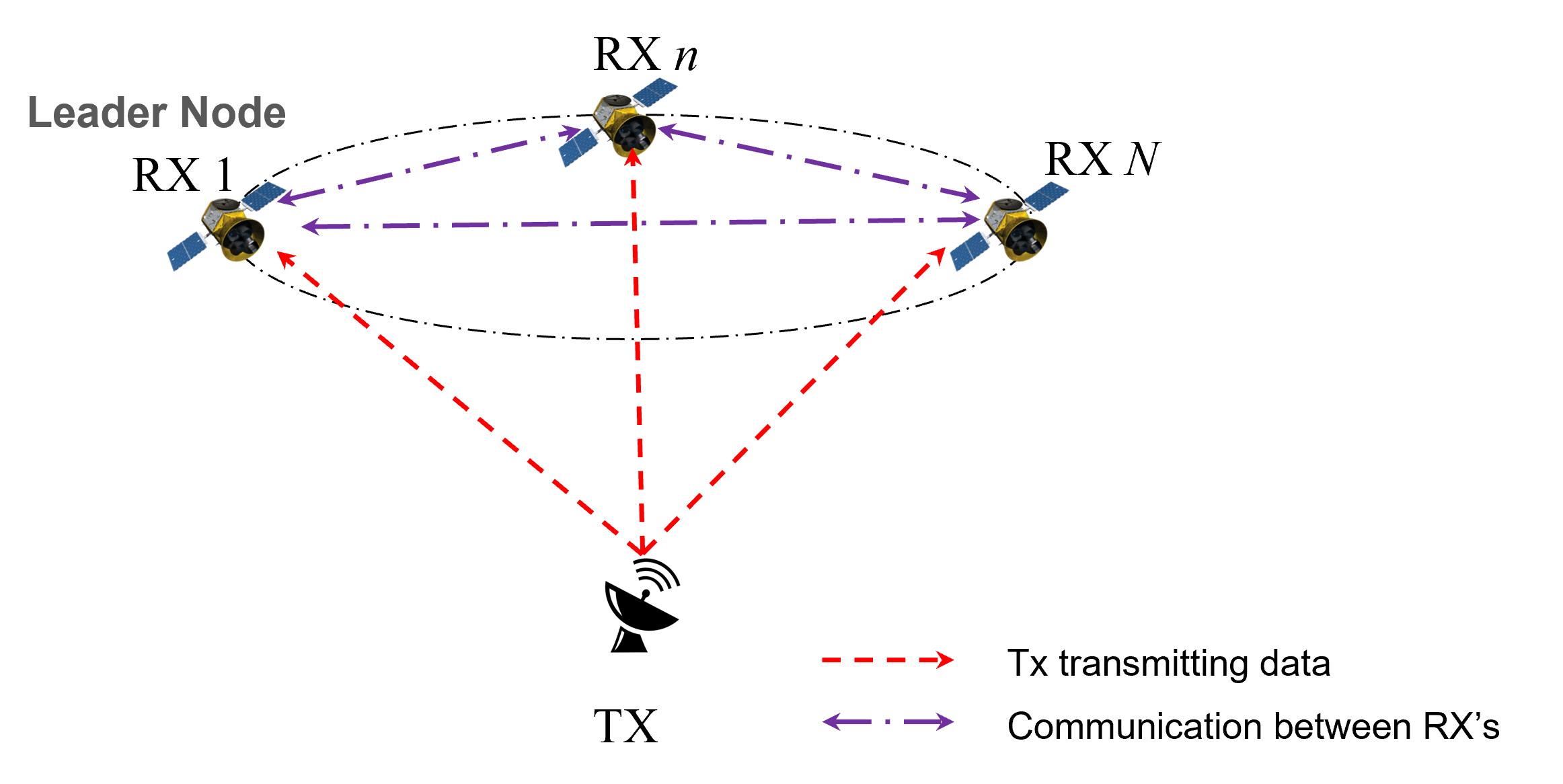}
    \captionsetup{justification=centering}
    \caption{Proposed RASHND architecture.}
    \label{fig:proposed_arch}
\end{figure}

Fig. \ref{fig:proposed_arch} illustrates the proposed system architecture, which consists of a SIMO setup with one transmitter (TX) and $N$ receivers (RX). 
% The TX represents a ground terminal that continuously transmits signals to the swarm of receivers in the air. 
The TX is a ground terminal that continuously transmits signals to the swarm of receivers in the air. 
% These RXs, which are part of a swarm system, are connected through an internal ad-hoc network that enables D2D communication and coordination.
These RXs, part of a swarm system, connect via an internal ad-hoc network for D2D communication and coordination. The RASHND network allows each device to share locally processed signal data with the rest of the swarm. A designated leader node within the swarm is responsible for collecting the processed data from all RX units and applying the appropriate combining algorithm to enhance signal quality. To ensure robustness and avoid single points of failure, the leader node role is dynamically assigned to different RXs sequentially. ]In the example shown in Fig. \ref{fig:proposed_arch}, RX$1$ is the leader node for that instance, demonstrating the flexibility and distributed control of the system.

\subsection{Experimental Testbed Setup}
\subsubsection{Indoor Testbed}
\begin{figure}
\centering
    %\includegraphics[width=0.8\columnwidth]{diagram1.png}
    % \includesvg[inkscapelatex=false,width=0.9\columnwidth]
    \includegraphics[width=0.85\columnwidth]{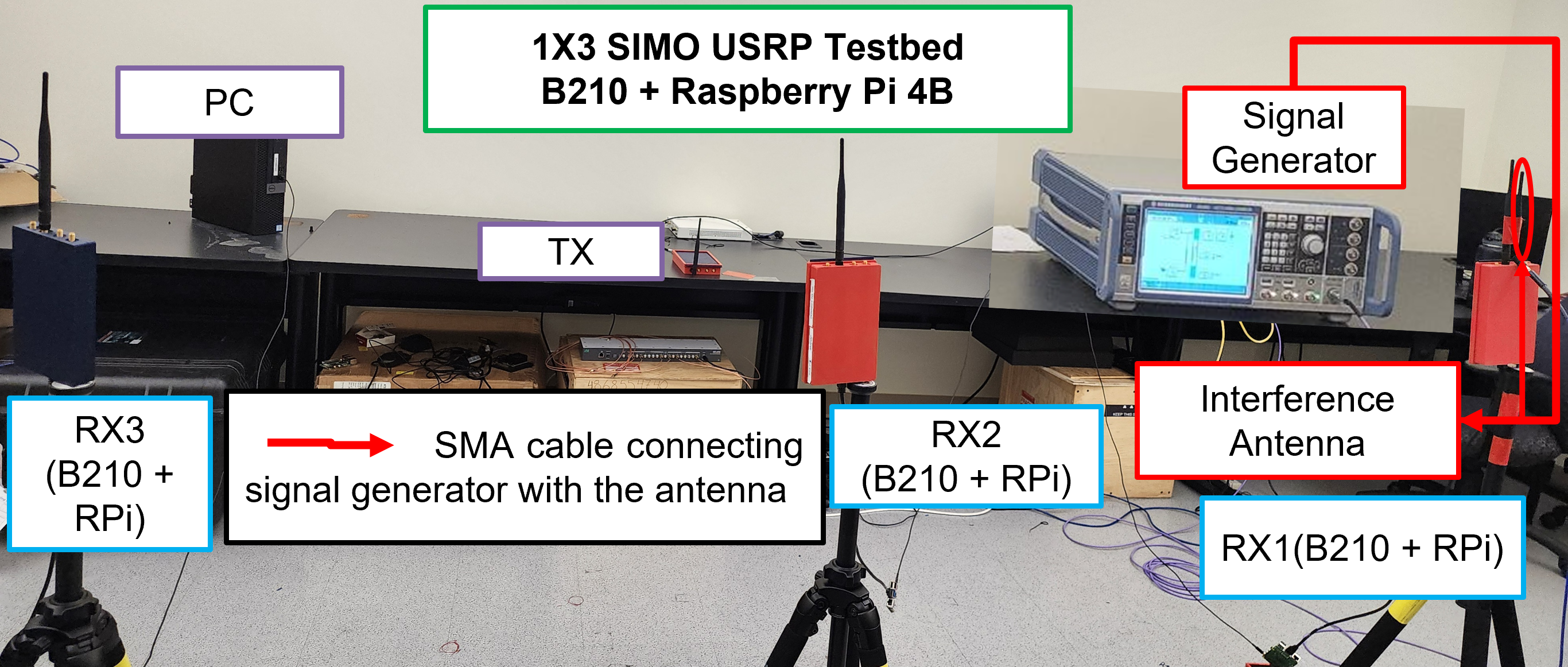}
    \captionsetup{justification=centering}
    \caption{Indoor testbed for $1\times3$ SIMO testbed.}
    \label{fig:indoor_testbed}
\end{figure}

Fig. \ref{fig:indoor_testbed} represents the testbed used to implement the system in an indoor setup. The setup includes three USRP B210s, each connected to a RPi, functioning as the RXs. An additional USRP B210 is connected to a PC and serves as the TX. The RPis control the RX units, and their compact, low-power design enables portable and flexible testing. An antenna connected to a signal generator is placed close to RX$1$ to introduce jamming into the system. 
% The USRP B210s support two wireless channels, allowing them to function as either single-antenna or multi-antenna devices depending on the use case. For this SIMO-based implementation, each USRP is configured to operate as a single-antenna radio.

\subsubsection{Outdoor Testbed}
\begin{figure}
\centering
    %\includegraphics[width=0.8\columnwidth]{diagram1.png}
    % \includesvg[inkscapelatex=false,width=0.9\columnwidth]
    \includegraphics[width=0.9\columnwidth]{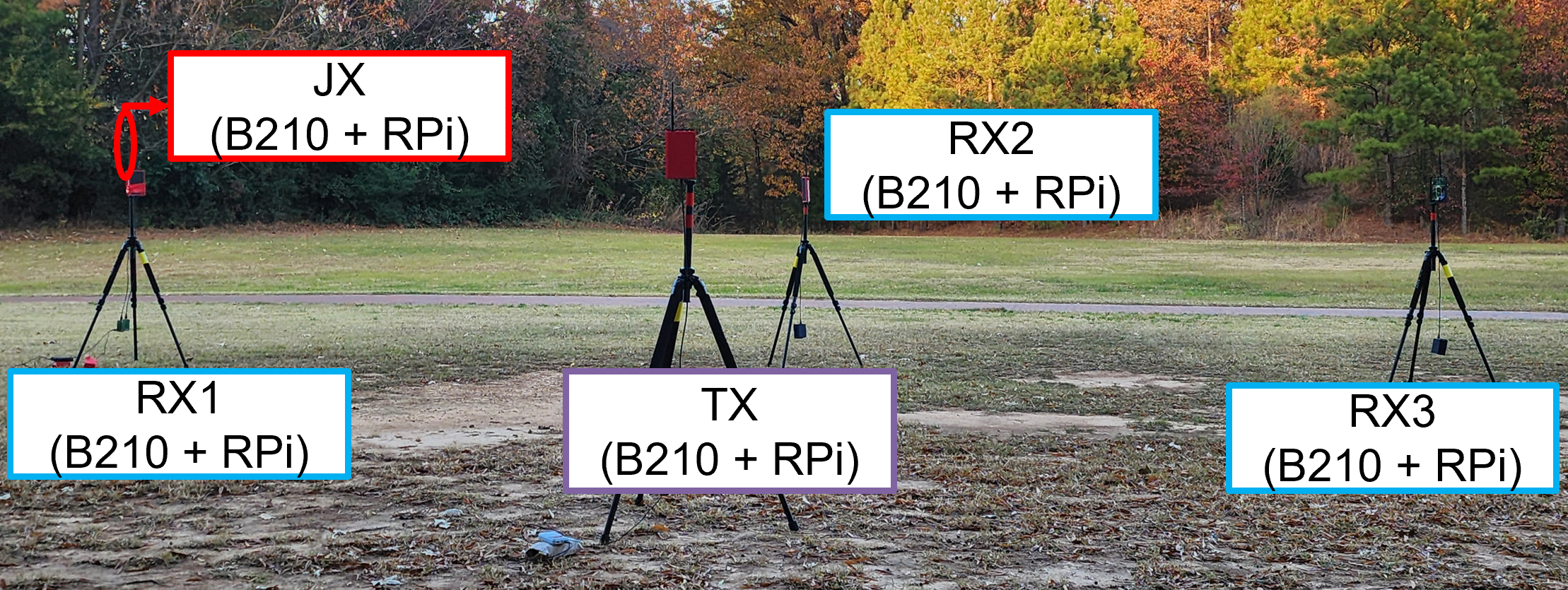}
    \captionsetup{justification=centering}
    \caption{Outdoor testbed for $1\times3$ SIMO testbed.}
    \label{fig:outdoor_testbed}
\end{figure}

Fig. \ref{fig:outdoor_testbed} represents the testbed used to implement the system in an outdoor environment. The setup includes five USRP B210s, each connected to a RPi, three of them functioning as RXs, with one each as TX and the jammer (JX). All our devices are placed on tripod stands and are distant from each other when compared to the indoor testbed. 

% Another USRP B210 is connected to an RPi and serves a Txs.  We also replace the signal generator with another USRP B210 connected to a RPi to serve as the jammer and is placed close to RX$1$. 

% \subsubsection{UAV Testbed}

\section{Self-healing Proposal}
This section presents the algorithms included in our setup and the proposed self-healing network.

\subsection{Algorithms Implemented in Self-Healing}
RASHND transitions among three algorithms to achieve superior signal quality, which are briefly described here.

\subsubsection{Distributed Maximal Ratio Combining (d-MRC)}
% The d-MRC algorithm uses a weighted combination of data sequences from multiple receivers based on individual channel quality \cite{milcom}.
The d-MRC algorithm combines data from multiple receivers weighted by channel quality \cite{milcom}.

The signal data at each of the RX can be written as,
\begin{align}
\mathbf{r}_i[k]= \sqrt{p_g}h_i\mathbf{x}[k] + \sqrt{p_j}h_{j,i}\mathbf{x_j}[k] + \mathbf{n}_i, 
\label{dt_rcvd_data_siso}
\end{align}
% where, $p_g$ and $p_j$ are the transmitter and jammer power respectively, $h_i$ and $h_{j,i}$ are SISO channel and the jammer channel information for the $i$th receiver, $\mathbf{x}$ and $\mathbf{x_j}$ are the input sequences for the transmitter and the jammer respectively, and $\mathbf{n}_i$ is the iid noise. 
where, $p_g$ and $p_j$ are the TX and JX power respectively, $h_i$ and $h_{j,i}$ are SISO channel and the jammer channel information for the $i$th receiver, $\mathbf{x}$ and $\mathbf{x_j}$ are the input sequences for the TX and the JX respectively, and $\mathbf{n}_i$ is the iid noise.

In accordance with Eq. \eqref{dt_rcvd_data_siso}, the data from all the receivers are collected into a single data structure for further processing. If there are a total of $N$ receivers, then $\mathbf{R} = [r_1,r_2,...,r_N]^T$, represents this data structure. d-MRC exploits uncorrelated channels to combine signals that carry same information. d-MRC multiplies the data at all receivers with a weight vector $\mathbf{u} =[u_1,u_2,...,u_N]^T$, such that the effective signal-to-noise ratio (SNR) is maximized. The interference is assumed to be negligible and thus, $\sqrt{p_j}h_{j,i}\mathbf{x_j}[k]$ can be ignored. The SNR for the combined result can thus be written as,
\begin{align}
\mathrm{SNR} = \frac{{p_g}{\sum^{N}_{(i=1)}{||u_i^Hh_i||^2}}}
{\sigma^2{\sum^{N}_{(i=1)}||u_i||^2}},
\label{mrc}
\end{align}
where $\sigma^2$ is the Additive White Gaussian Noise (AWGN) noise power level, $\mathbf{h} = [h_1,h_2,...,h_N]^T$ represents a vector of channel state information from $N$ receivers and $(.)^H$ is the Hermitian operator.

By optimizing as in \cite{heath2017introduction} and applying the Cauchy-Schwarz Inequality, we obtain the d-MRC weight $\mathbf{u}_{d-MRC}$ as,
\begin{align}
\mathbf{u}_{\mathrm{d-MRC}} = \mathbf{h}^H.
\end{align}

\subsubsection{Distributed Linear Minimum Mean Square Error Estimation (d-LMMSE)}
% The d-LMMSE algorithm also utilizes a weighted combination of data sequence but includes the interference data and thus, is useful for signals with high levels of interference such as jamming \cite{milcom}.
The d-LMMSE algorithm uses a weighted combination of data, including interference, making it effective for signals with high interference such as jamming \cite{milcom}. The effective signal-to-interference-plus-noise ratio (SINR) can be written as,
\begin{align}
\mathrm{SINR} = \frac{{p_g}{\sum^{N}_{(i=1)}{||u_i^Hh_i||^2}}}
{{\sigma^2\sum^{N}_{(i=1)}||u_i^H ||^2} +{{p_j}\sum^{N}_{(i=1)}||u_i^Hh_{j,i} ||^2}}.
\end{align}
% An optimal value of $\mathbf{u}$ is obtained by maximizing SINR by solving the above equation as a minimum mean squared error optimization problem. By assuming that the transmitted signals are legitimate and thus, external interference signals and additive noise are mutually independent, we obtain, 
The optimal value of $\mathbf{u}$ is found by maximizing SINR via the above equation as a minimum mean squared error problem. Assuming transmitted signals are legitimate, so that external interference and noise are independent, we obtain,
\begin{align}
\mathbf{u}_{d-LMMSE} = {p_g}[{p_g}\mathbf{h}\mathbf{h}^H + {p_j}\mathbf{h}_j\mathbf{h}_j^H + \sigma^2\mathbf{I}]^{-1}\mathbf{h},
\end{align}
where, $\mathbf{I}$ is the identity matrix of size $N \times N$.

\subsubsection{Selection Combining (SC)}
% Selection Combining selects the best among the $N$ receivers at any given point of time \cite{seldiv}.
Selection Combining selects the best of the $N$ receivers at any time \cite{seldiv}.
Mathematically, this is similar to Eq. \eqref{mrc}, where RXs heavily affected by interference have a weight vector of 0.
% Mathematically, this will be similar to Eq. \eqref{mrc}, where the RX's that are highly impacted with interference will have a weight vector of $0$. 
This leaves the combined signal to be equal to the signal with legitimate data, thus making the weight vector for that particular RX to be $1$.

% \subsection{Results for Standalone Algorithm Implementation}
%%%%why use mrc when lmmse exists? mrc is computatinally easier plus doesnt need jamming data. we show both here for reference but in reality we can save jamming data information and computation power.

\subsection{State Machine Model for Self-Healing}
\begin{figure}
\centering
    %\includegraphics[width=0.8\columnwidth]{diagram1.png}
    % \includesvg[inkscapelatex=false,width=0.65\columnwidth]{images/state_machine.drawio.svg}
    \includegraphics[width=0.65\columnwidth]{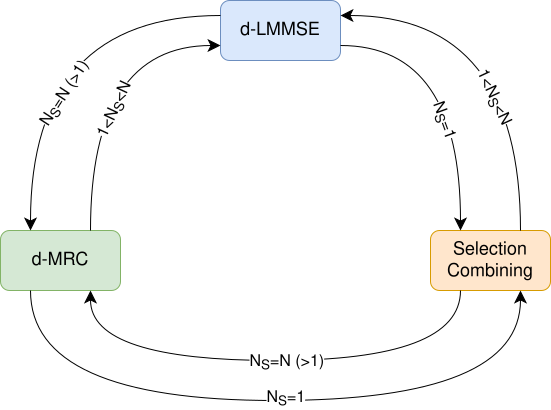}
    \captionsetup{justification=centering}
    \caption{State machine for self-healing.}
    \label{fig:state_machine_phy}
\end{figure}

Self-healing is implemented to make the system independent and have the best possible combining algorithm implemented at all times in order to obtain a superior quality signal. Fig. \ref{fig:state_machine_phy} represents the state machine used to implement the combining method. The SHN algorithm is selected based on the number of receivers, $Ns$, having bit error rates (BER) greater than the threshold. One among d-MRC, d-LMMSE and SC gets chosen based on the BER values. 
% N receivers listed in the sentence above.
% $N$ represents the total number of RXs.

Fig. \ref{fig:state_machine_phy} describes the implementation of the three combining algorithms. When all the RXs receive high quality signals, the BER for each of the RX units will be greater than the given threshold. Here, the number of receivers with BER greater than the threshold, $N_S$, will be equal to the total number of available receivers, i.e, $N_S=N$, prompting the system to implement d-MRC. In cases of extremely high interference, where only one of the RXs receive a good signal, i.e, $N_S=1$, the system implements SC, where the best performing receiver is chosen. 
Another scenario is when the number of RXs receiving high quality signal is more than one but less than the total number of available receivers.
% Another scenario is when more than one, but not all, RXs receive a high-quality signal. Here, the BER is greater than the threshold for $1<N_S<N$ and thus d-LMMSE gets implemented.

\section{Inter-Device Links}
To develop a self-sufficient swarm system with IDL capabilities, we designed a network of RXs capable of direct D2D communication. Each RX shares data over a dedicated wireless channel, different from the SIMO communication channel. The RASHND architecture allows the RXs to perform signal combining locally, removing the need for an external computer and making the swarm system autonomous.

\begin{figure}
\centering 
% \includesvg[inkscapelatex=false,width=0.4\columnwidth]
\includegraphics[width=0.4\columnwidth]{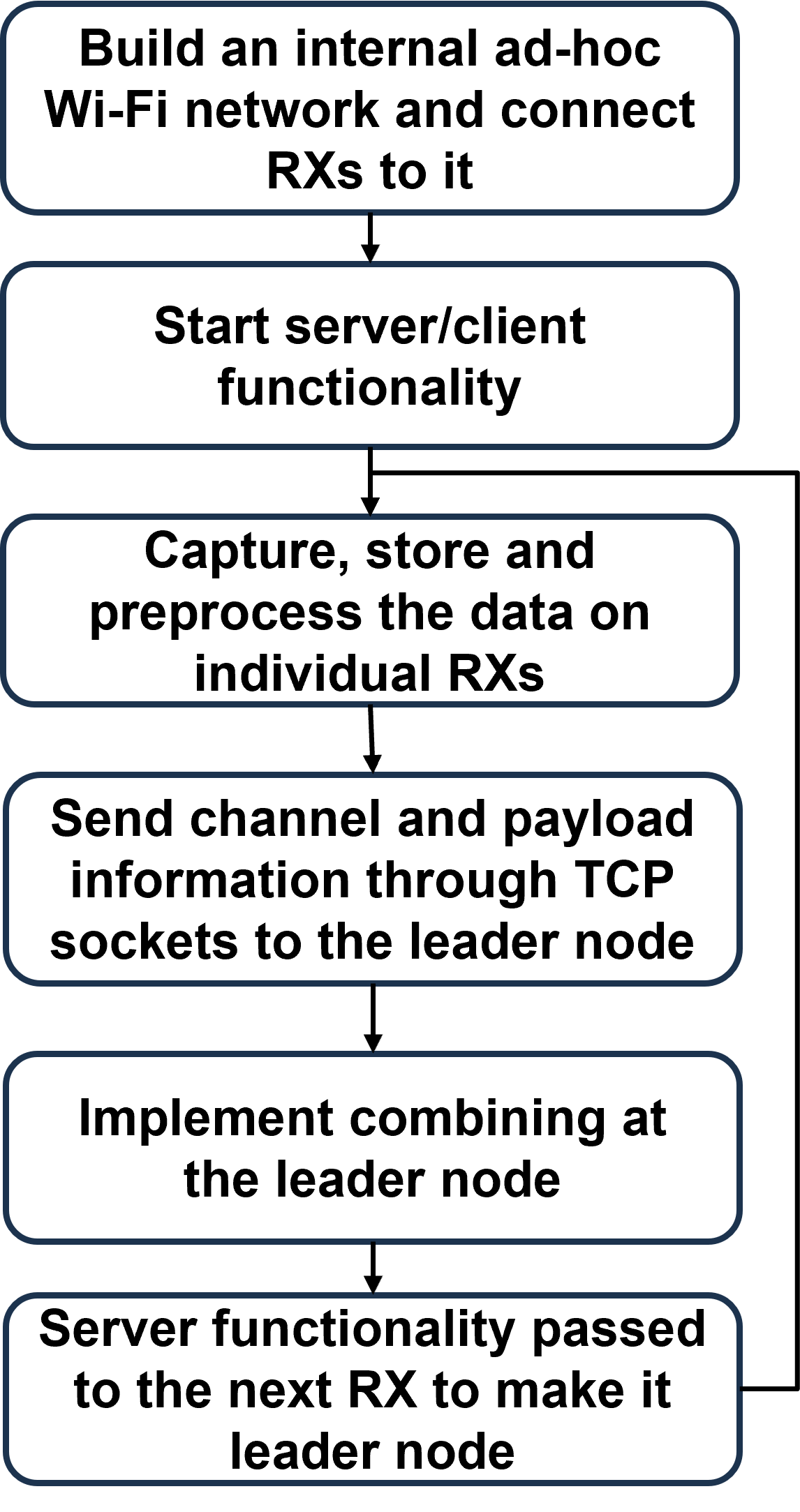}
    \captionsetup{justification=centering}
    \caption{Flow diagram for inter-device links.}
    \label{fig:line_diagram}
\end{figure}

% D2D communication is implemented using TCP sockets, where each RX can function both as a server and client.
TCP sockets enable D2D communication, allowing each RX to act as both client and server. The flow diagram in Fig. \ref{fig:line_diagram} represents the system flow. Upon startup, all devices connect to a common Wi-Fi network using the onboard network interface cards on the RPis. Once connected, the RXs establish TCP server-client connections in a dynamic, rotating fashion \cite{dyspan}. 
% Initially, RX$1$ assumes the role of the server, acting as the leader node, while the other RXs connect as clients. After a processing cycle, the server responsibility shifts to RX$2$, then to RX$3$, and so on, ensuring that each device takes turns coordinating the data exchange and processing.
Initially, RX$1$ acts as the server (leader), with other RXs as clients. After each processing cycle, the server role rotates to RX$2$, RX$3$, and so on, ensuring all devices take turns coordinating data exchange and processing.

% During operation, each RX captures and processes its own signal data, and stores both channel and payload information locally. The client RXs then transmit their processed data to the current leader node via the established TCP sockets.  At the leader node, the system computes the BER for each receiver and compares these values to a predefined threshold. Based on the BER analysis, the leader node selects and implements the most suitable combining algorithm to optimize performance. Once the combining process is complete, the leader node capabilities are passed to the next RX in the sequence, and the process repeats, enabling continuous and distributed operation without external oversight.
During operation, each RX captures and processes its signal data, storing channel and payload information locally. Client RXs then transmit the processed data to the current leader node via established TCP sockets. The leader node computes each receiver’s BER, compares it to a threshold, and selects the optimal combining algorithm. Once the combining process is complete, the leader node capabilities are passed to the next RX in the sequence, and the process repeats, enabling continuous and distributed operation without external oversight.

To enhance robustness, the RXs are programmed to automatically detect and recover from communication failures. If the connection is lost, the devices attempt to reconnect without requiring the system to restart. Should a particular RX become unavailable, the system simply excludes it from the combining process and continues operating with the remaining devices. This fault-tolerant design ensures the UAV swarm remains resilient and independent of any single node.

\section{Over the air Implementation}
This section discusses the results obtained for self-healing network in the indoor and outdoor testbed. We present the results for bitstream and image data, along with the results for UAV based testing using the AERPAW platform.
% We present the results for bitstream and image data. We also present data obtained from UAV based testing using the AERPAW platform.

\subsection{Self-Healing on Indoor Testbed}
Our algorithms are at first validated in our indoor testbeds. 
% The results for the same are discussed below.

\subsubsection{Results for Bitstream Data}
The system is first tested for bitstream data, where a random set of data is generated and modulated using QPSK modulation and then, appended with Zadoff-Chu preamble sequence. We use a center frequency of $2.55$ GHz, a bandwidth of $1$ MHz and RX power of $-49$ dBm. The data is continuously transmitted at a constant power of $-4$ dBm. The entire reception is automated to work with self-healing. Each B210 captures the data, which the RPi stores and compensates for timing, frequency and phase difference, determines the channel estimates and the payload information. The channel estimates and payload information are shared with the leader node through D2D communication to implement the combining algorithm and obtain the data rates.
% The data is continuously transmitted from the transmitter at a constant power of $−4$ dBm.
\begin{figure}
\centering
    % \includesvg[inkscapelatex=false,width=0.75\columnwidth]{images/bitstream_inlab.png}
    % \includesvg[inkscapelatex=false,width=0.7\columnwidth]
    \includegraphics[width=0.7\columnwidth]{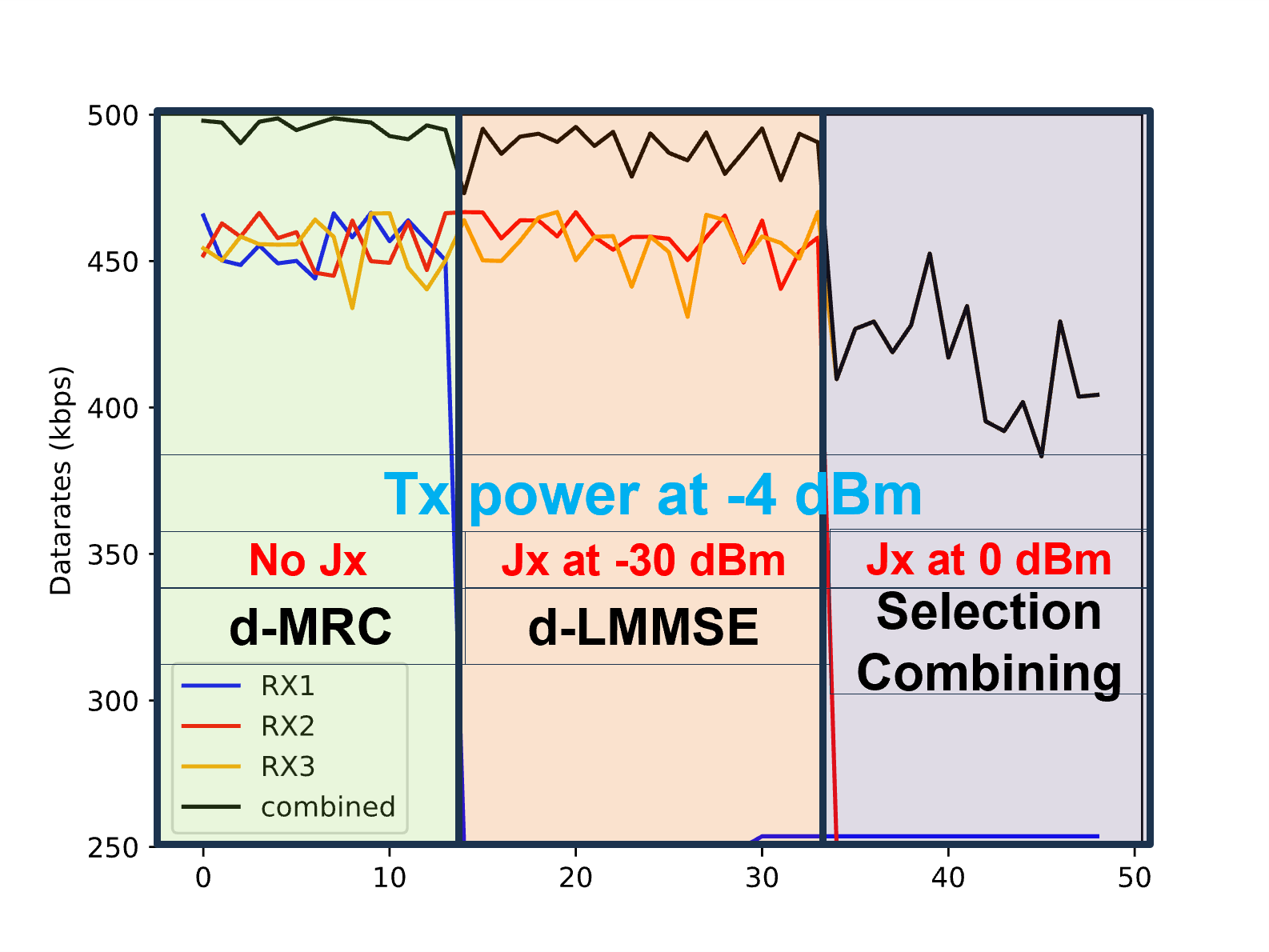}
    \captionsetup{justification=centering}
    \caption{Data rates for bitstream data for indoor testbed.}
    \label{fig:bitstream_indoor}
\end{figure}

Fig. \ref{fig:bitstream_indoor} represents the continuous plot for bitstream data through self-healing implemented using the automated setup. 
% Signals are transmitted at a constant power of $-4$ dBm. We start our experiments with no interference and observe that all three receivers have similar performance and have high data rates, suggesting that the d-MRC algorithm was implemented since the BER for all receivers exceeded the threshold, as specified by the state machine.
% Signals are transmitted at a constant power of $-4$ dBm. 
% With no interference, all three receivers perform similarly, achieving high data rates, suggesting that the d-MRC algorithm was implemented since the BER for all receivers exceeded the threshold. 
% We then introduce external interference with power $-30$ dBm and notice that the receiver closest to the interference antenna, i.e, RX$1$ is impacted and thus, d-LMMSE is implemented. 
With no interference, all three receivers achieve high data rates, showing that the d-MRC algorithm is applied since their BERs exceed the threshold.
We then introduce $-30$ dBm external interference. The receiver nearest to the interference antenna, RX$1$, is affected, so d-LMMSE is applied. Further increasing the interference power to $0$ dBm impacts both RX$1$ and RX$2$. Thus, SC is implemented, and the combined data rate equals the best performing receiver, RX$3$.

\subsubsection{Results for Image Data}
% Since self-healing performs well on bitstream data, it is extended to larger datasets by applying it to image data.
% A small airplane image is converted to grayscale and vectorized, then converted into bitstream before transmitting it, similar to the bitstream data transmission. The data received is processed similar to the bitstream data and the final images are reconstructed and displayed along with the data rates. 
Since self-healing works well on bitstream data, it is extended to larger datasets like images. A small airplane image is converted to grayscale and vectorized, then turned into bitstream for transmission, similar to previous bitstream data. The received data is processed similar to the bitstream data, and the reconstructed images are displayed along with their data rates.

\begin{figure}
\centering
    % \includesvg[inkscapelatex=false,width=0.75\columnwidth]{images/images_inlab.svg}
    % \includesvg[inkscapelatex=false,width=0.7\columnwidth]
    \includegraphics[width=0.7\columnwidth]{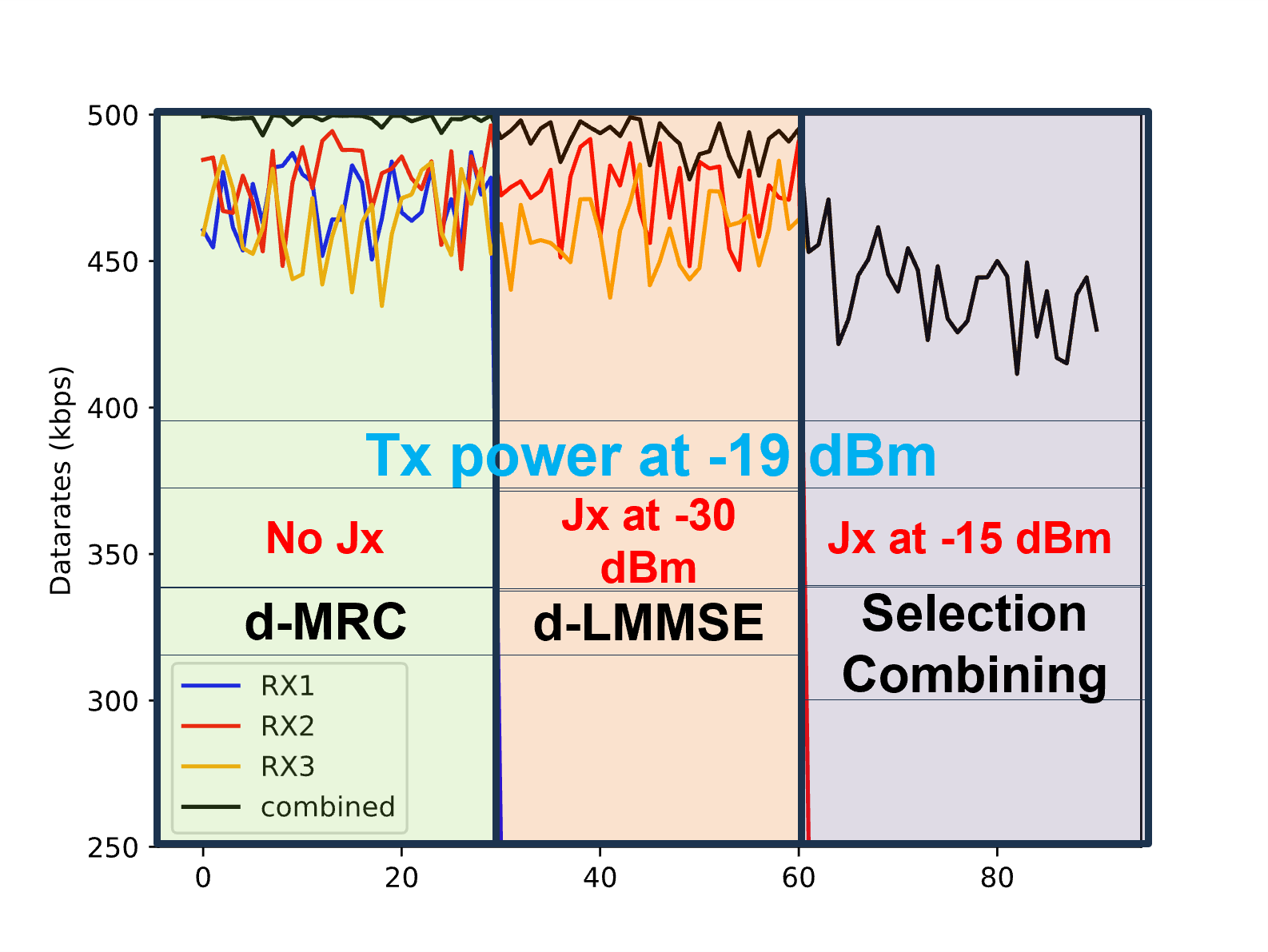}
    % \includesvg[inkscapelatex=false,width=0.8\columnwidth]{images/images_inlab.svg}
    \captionsetup{justification=centering}
    \caption{Data rates for images data for indoor testbed.}
    \label{fig:images_indoor}
\end{figure}

Fig. \ref{fig:images_indoor} represents the output for image transmission. 
% We see that the combined data rate is always better than the individual receivers, demonstrating our algorithms work well even for larger datasets like images. 
The combined data rate consistently exceeds that of individual receivers, showing our algorithms perform well even on larger datasets like images. The displayed images also demonstrate that not only is the data rate improved, but the overall quality of the received image is enhanced as well. 
 
We transmit data at a constant power of $-19$ dBm. With no interference, all three receivers have similar performance along with high data rates, suggesting that d-MRC algorithm was implemented. We then introduce an external interference with power $-30$ dBm and notice that the receiver closest to the interference antenna, RX$1$ is impacted and doesn't receive good data, thus, d-LMMSE is implemented. We further increase the interference power to $-15$ dBm and notice that the two receivers closer to the external interference antenna are heavily impacted, leaving only RX$3$ to receive valid data and have higher BER than the threshold, thus, SC gets implemented with data rate matching the data rate of RX$3$. 

\begin{figure*}
\centering
     % \begin{subfigure}{0.29\textwidth}
     \begin{subfigure}{0.27\textwidth}
    \centering
      % \includesvg[inkscapelatex=false,width=\textwidth]
      \includegraphics[width=\textwidth]{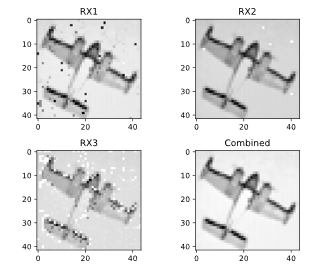}
    \captionsetup{justification=centering}
    \caption{Image for d-MRC Combining}
    \label{fig:mrc_indoor}
    \end{subfigure}
    \centering
    % \begin{subfigure}{0.29\textwidth}
    \begin{subfigure}{0.27\textwidth}
    \centering
    % \includesvg[inkscapelatex=false,width=\textwidth]
    \includegraphics[width=\textwidth]{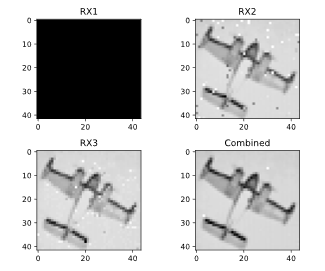}
    \captionsetup{justification=centering}
    \caption{Image for d-Lmmse Combining}
    \label{fig:lmmse_indoor}
    \end{subfigure}
      % \begin{subfigure}{0.29\textwidth}
      \begin{subfigure}{0.27\textwidth}
    \centering
    % \includesvg[inkscapelatex=false,width=\textwidth]
    \includegraphics[width=\textwidth]{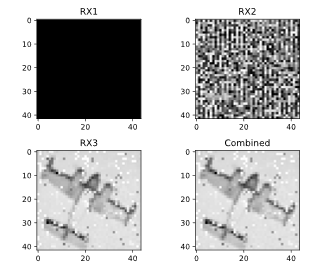}
    \captionsetup{justification=centering}
    \caption{Image for Selection Combining}
    \label{fig:siso_indoor}
    
    \end{subfigure}
     \captionsetup{justification=centering}
    \caption{Received images for indoor testbed}
\end{figure*}

Fig. \ref{fig:mrc_indoor} - Fig. \ref{fig:siso_indoor} are the sample reconstructed images at the receivers, displayed for each of the three cases. The figures clearly illustrate the RASHND resilience. Fig. \ref{fig:mrc_indoor} is the output for d-MRC implementation with all the RXs receiving high quality data. Fig. \ref{fig:lmmse_indoor} represents d-LMMSE implementation and shows that RX$1$ is highly impacted, with no useful information and the combined image is the improvised version of RX$2$ and RX$3$. Fig. \ref{fig:siso_indoor} illustrates the output for SC with RX$1$ and RX$2$ being heavily impacted. The combined image utilizes the best RX, which here is RX$3$.

\subsection{Self-Healing on Outdoor Testbed}
We further verify the algorithms through outdoor field tests.
% To further verify our algorithms, we conduct outdoor field tests and the results are presented in Figs. \ref{fig:bitstream_outdoor} - \ref{fig:outdoor_full}.
% for the same are discussed here.

\subsubsection{Results for Bitstream Data}
\begin{figure}
\centering
    % \includesvg[inkscapelatex=false,width=0.75\columnwidth]{images/bitstream_outdoor.svg}
    % \includesvg[inkscapelatex=false,width=0.7\columnwidth]
    \includegraphics[width=0.7\columnwidth]{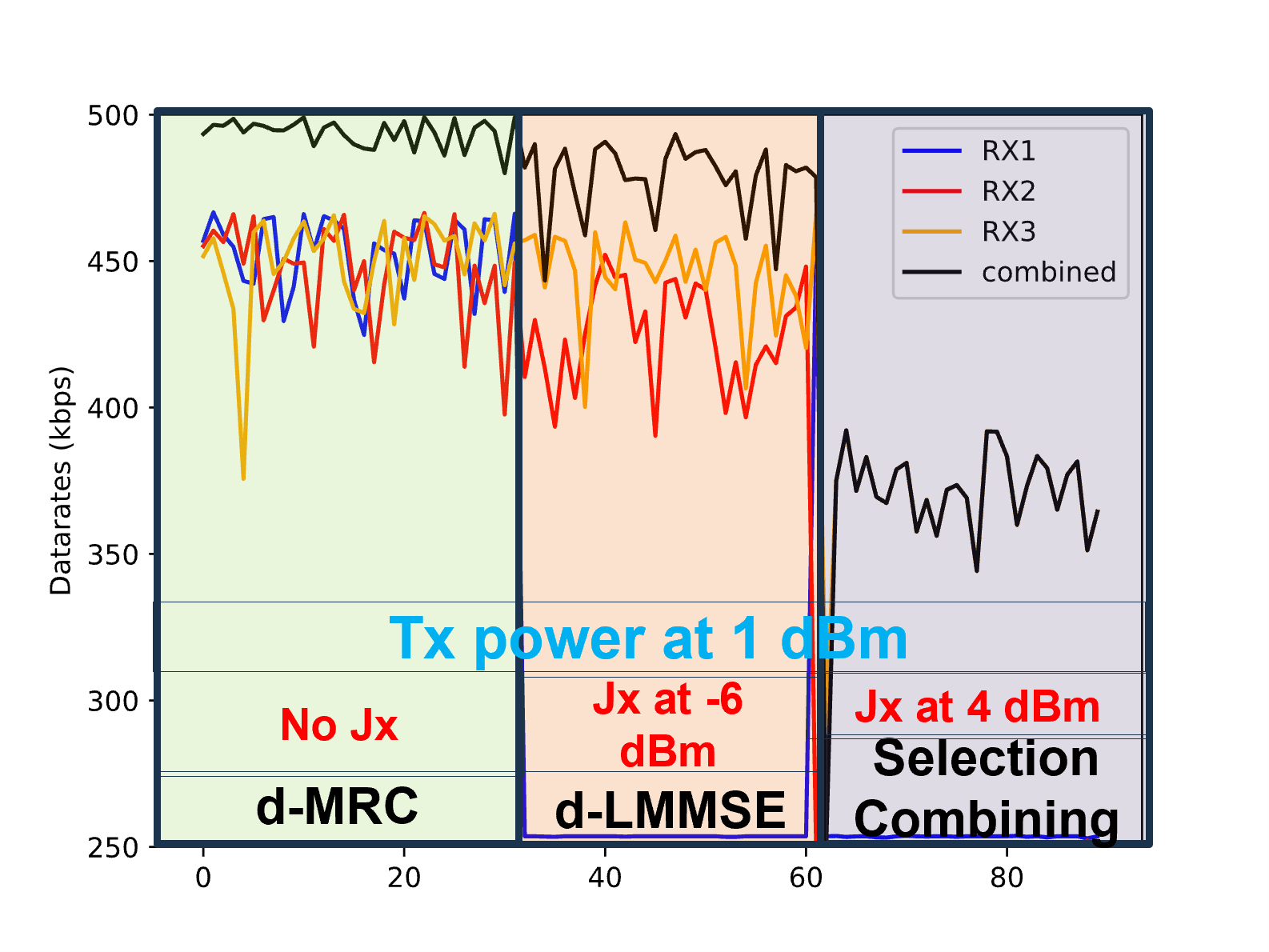}
    % \includesvg[inkscapelatex=false,width=0.8\columnwidth]{images/bitstream_outdoor.svg}
    \captionsetup{justification=centering}
    \caption{Data rates for bitstream data for outdoor testbed.}
    \label{fig:bitstream_outdoor}
\end{figure}

Similar to the indoor testbed, we first perform experiments for bitstream data by continuously transmitting with a transmit power of $1$ dBm. We start with no interference case where d-MRC algorithm gets implemented. 
% We then introduce a constant interference of $-6$ dBm and notice that the receiver closest to the interference antenna, i.e., RX$1$ is heavily impacted and thus d-LMMSE gets implemented. 
We then introduce a constant $-6$ dBm interference, which impacts RX$1$ and thus d-LMMSE is implemented.
A further increase in the interference level to $4$ dBm impacts both RX$1$ and RX$2$ and thus SC is implemented with the data rates being equal to that of the best RX, which here is RX$3$ as shown in Fig. \ref{fig:bitstream_outdoor}.

\begin{figure}
\centering
    % \includesvg[inkscapelatex=false,width=0.75\columnwidth]{images/images_outdoor.svg}
    % \includesvg[inkscapelatex=false,width=0.7\columnwidth]
    \includegraphics[width=0.7\columnwidth]{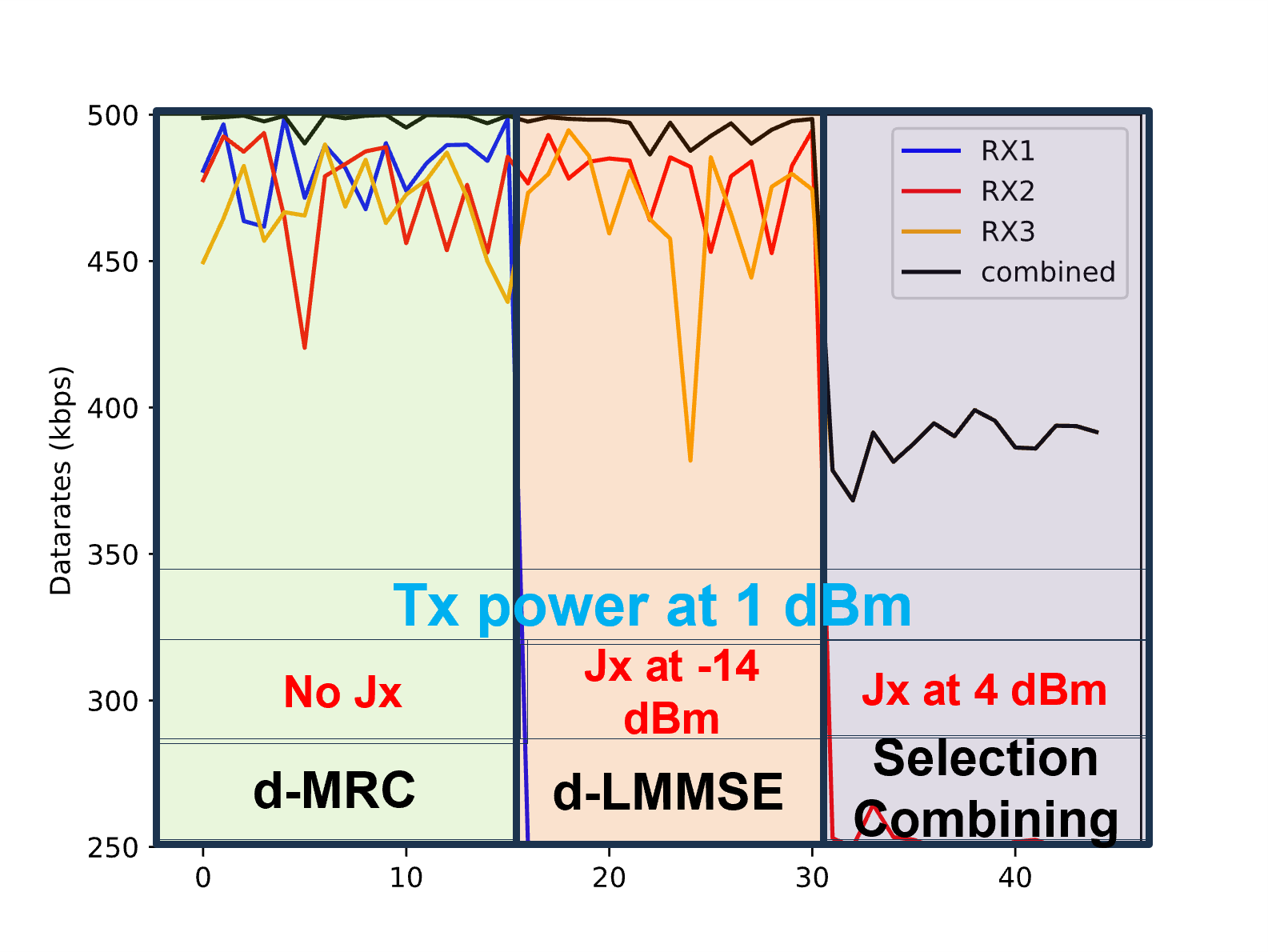}
    % \includesvg[inkscapelatex=false,width=0.8\columnwidth]{images/images_outdoor.svg}
    \captionsetup{justification=centering}
    \caption{Data rates for images data for outdoor testbed.}
    \label{fig:images_outdoor}
\end{figure}

% \begin{figure}
% \centering
%     % \includegraphics[width=0.8\columnwidth]{diagram1.png}
%     \includesvg[inkscapelatex=false,width=0.60\columnwidth]{images/aerpaw_path1.svg}
%     \captionsetup{justification=centering}
%     \caption{Path for AERPAW experiments.}
%     \label{fig:aerpaw_path}
% \end{figure}

\begin{figure*}
\centering
     % \begin{subfigure}{0.29\textwidth}
     \begin{subfigure}{0.27\textwidth}
    \centering
      % \includesvg[inkscapelatex=false,width=\textwidth]
      \includegraphics[width=\textwidth]{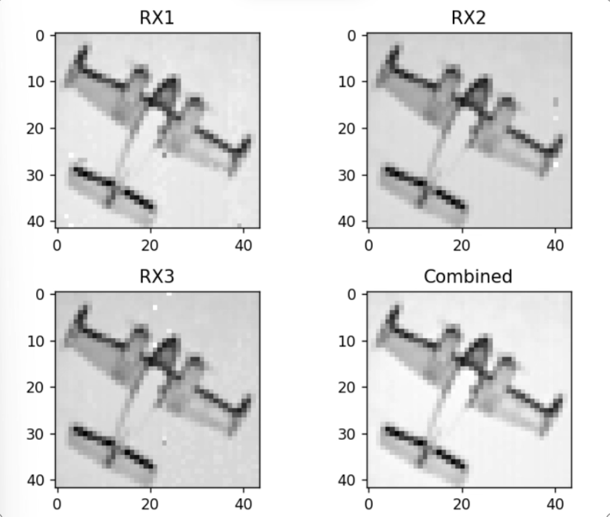}
    \captionsetup{justification=centering}
    \caption{Image for d-MRC Combining}
    \label{fig:mrc_outdoor}
    \end{subfigure}
    \centering
    % \begin{subfigure}{0.29\textwidth}
    \begin{subfigure}{0.27\textwidth}
    \centering
    % \includesvg[inkscapelatex=false,width=\textwidth]
    \includegraphics[width=\textwidth]{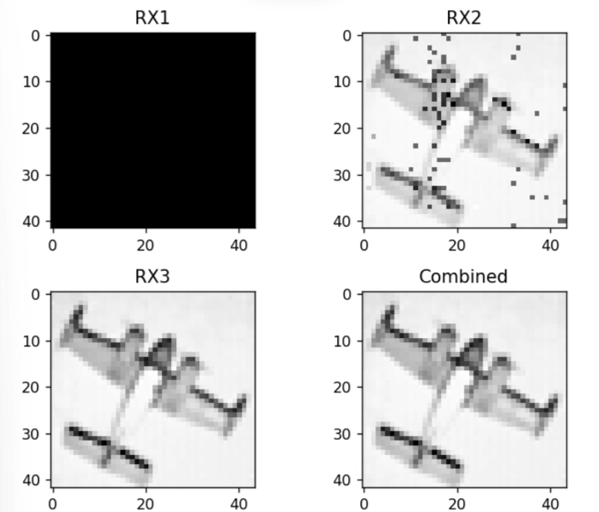}
    \captionsetup{justification=centering}
    \caption{Image for d-LMMSE Combining}
    \label{fig:lmmse_outdoor}
    \end{subfigure}
      % \begin{subfigure}{0.29\textwidth}
      \begin{subfigure}{0.28\textwidth}
    \centering
    % \includesvg[inkscapelatex=false,width=\textwidth]
    \includegraphics[width=\textwidth]{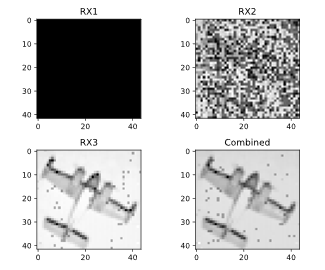}
    \captionsetup{justification=centering}
    \caption{Image for Selection Combining}
    \label{fig:siso_outdoor}
    
    \end{subfigure}
     \captionsetup{justification=centering}
    \caption{Received images for outdoor testbed.}
    \label{fig:outdoor_full}
\end{figure*}

\subsubsection{Results for Image Data}
Outdoor experiments for images are performed same as before where data is continuously transmitted with a transmit power of $1$ dBm. We begin with no interference case where d-MRC algorithm gets implemented. Introducing a constant interference of $-14$ dBm heavily impacts RX$1$ and thus d-LMMSE gets implemented.
An increase in the interference level to $4$ dBm impacts both RX$1$ and RX$2$, leading to the implementation of SC. Fig. \ref{fig:images_outdoor} represents the continuous plot for image data through self-healing.

Fig. \ref{fig:mrc_outdoor} - Fig. \ref{fig:siso_outdoor} are the sample reconstructed images at the receivers, displayed for each of the three cases. The results represent a robust prototype system.
% The figures clearly illustrate the working of our algorithms. 
Fig. \ref{fig:mrc_outdoor} is the output for d-MRC implementation with all the RXs receiving high quality data.  Fig. \ref{fig:lmmse_outdoor} represents d-LMMSE implementation, clearly showing RX$1$ is highly impacted and the combined image is the improvised version of RX$2$ and RX$3$.  Fig. \ref{fig:siso_outdoor} shows SC where RX$1$ and RX$2$ are heavily impacted. The combined image selects the best RX, which here is RX$3$.

% \begin{figure}
% \centering
%     % \includegraphics[width=0.8\columnwidth]{diagram1.png}
%     \includesvg[inkscapelatex=false,width=0.65\columnwidth]{images/aerpaw_path1.svg}
%     \captionsetup{justification=centering}
%     \caption{Path for AERPAW experiments.}
%     \label{fig:aerpaw_path}
% \end{figure}

\begin{figure*}
\centering
     % \begin{subfigure}{0.29\textwidth}
     \begin{subfigure}{0.27\textwidth}
    \centering
      % \includesvg[inkscapelatex=false,width=\textwidth]
      \includegraphics[width=\textwidth]{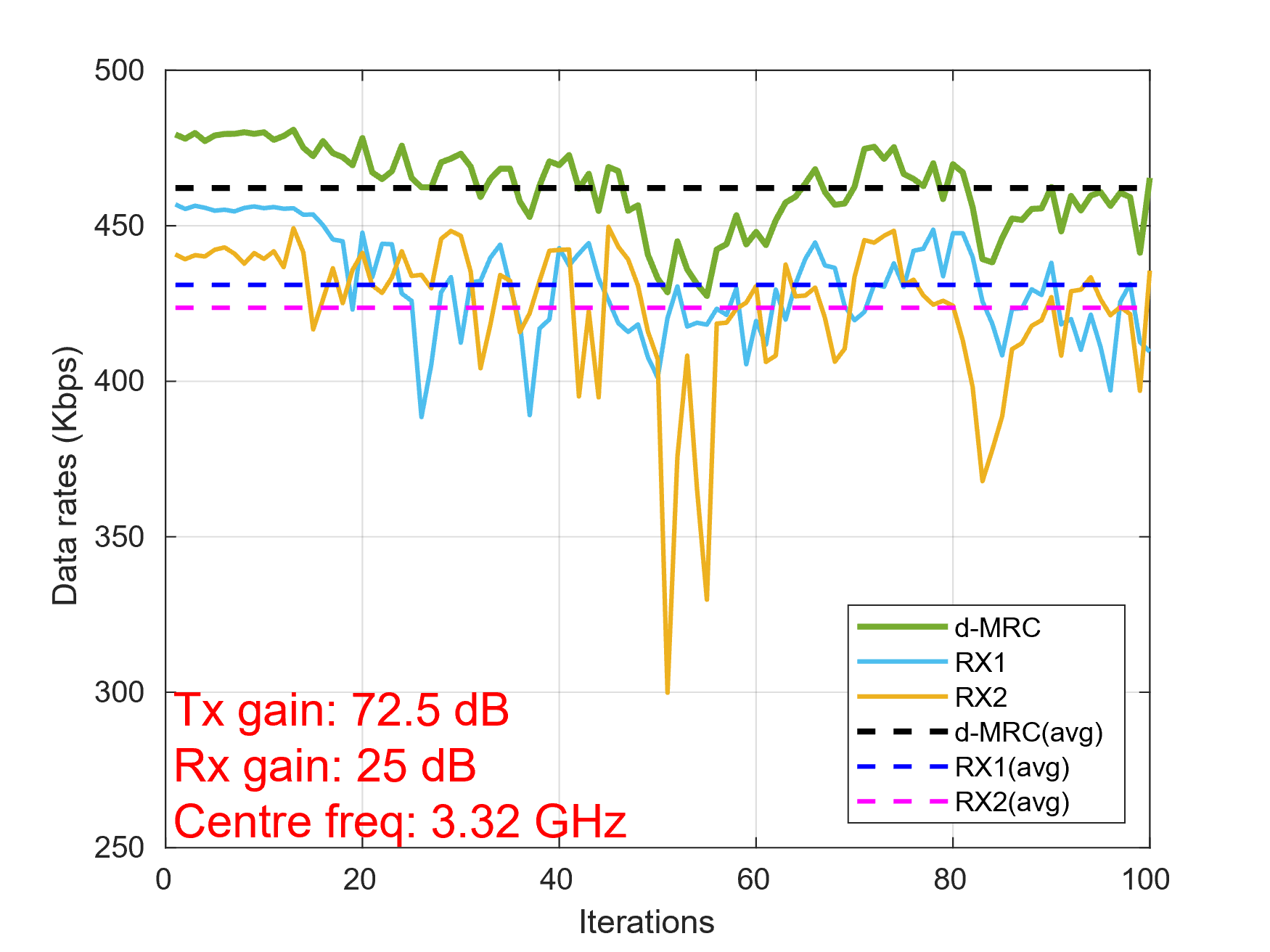}
    \captionsetup{justification=centering}
    \caption{UAVs taking different paths}
    \label{fig:Datarate - UAVs taking different paths}
    \end{subfigure}
    \centering
    % \begin{subfigure}{0.29\textwidth}
    \begin{subfigure}{0.27\textwidth}
    \centering
    % \includesvg[inkscapelatex=false,width=\textwidth]
    \includegraphics[width=\textwidth]{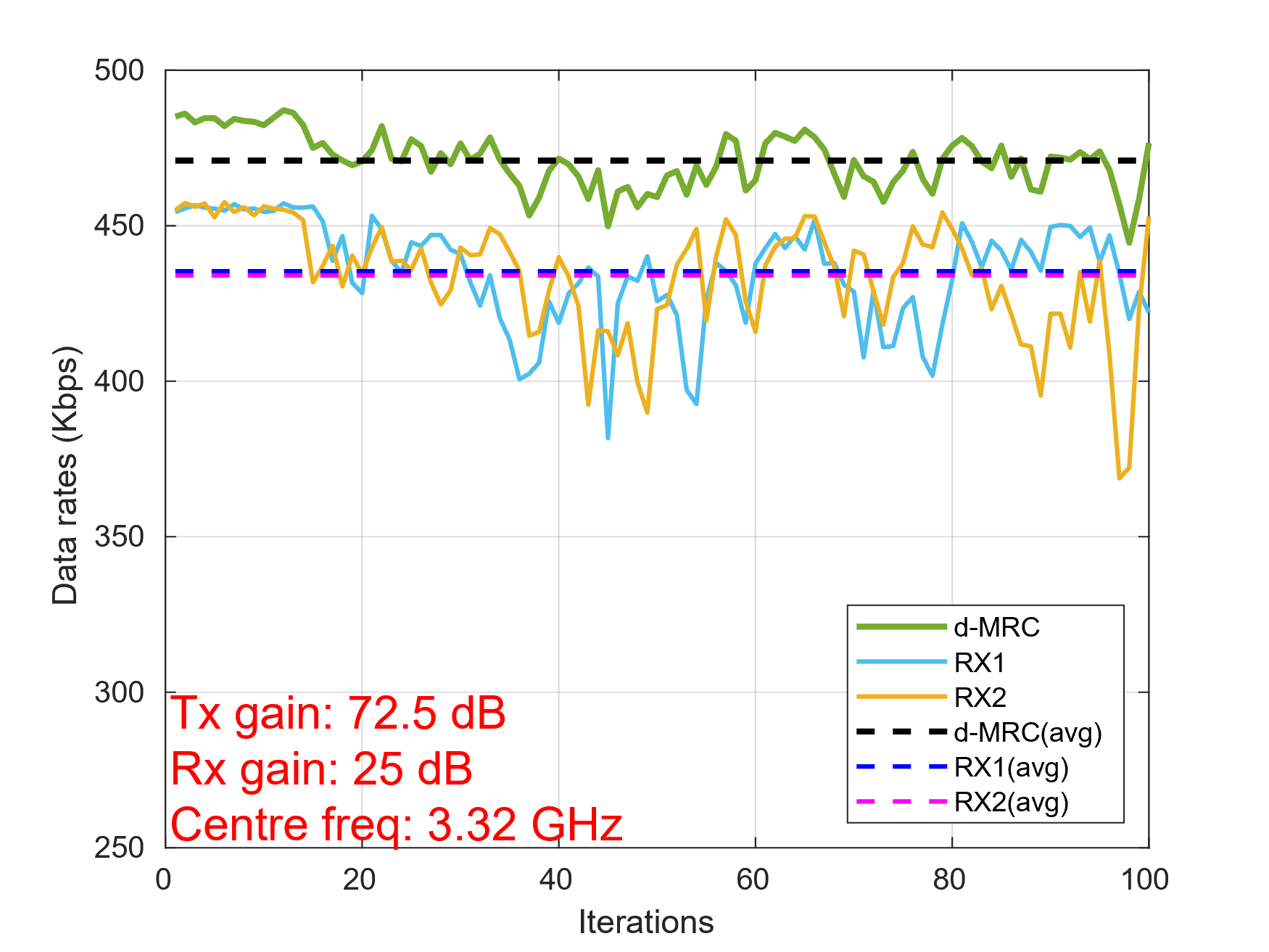}
    \captionsetup{justification=centering}
    \caption{Both UAVs taking path1}
    \label{fig:Datarate - Both UAVs taking path1}
    \end{subfigure}
      % \begin{subfigure}{0.29\textwidth}
      \begin{subfigure}{0.27\textwidth}
    \centering
    % \includesvg[inkscapelatex=false,width=\textwidth]
    \includegraphics[width=\textwidth]{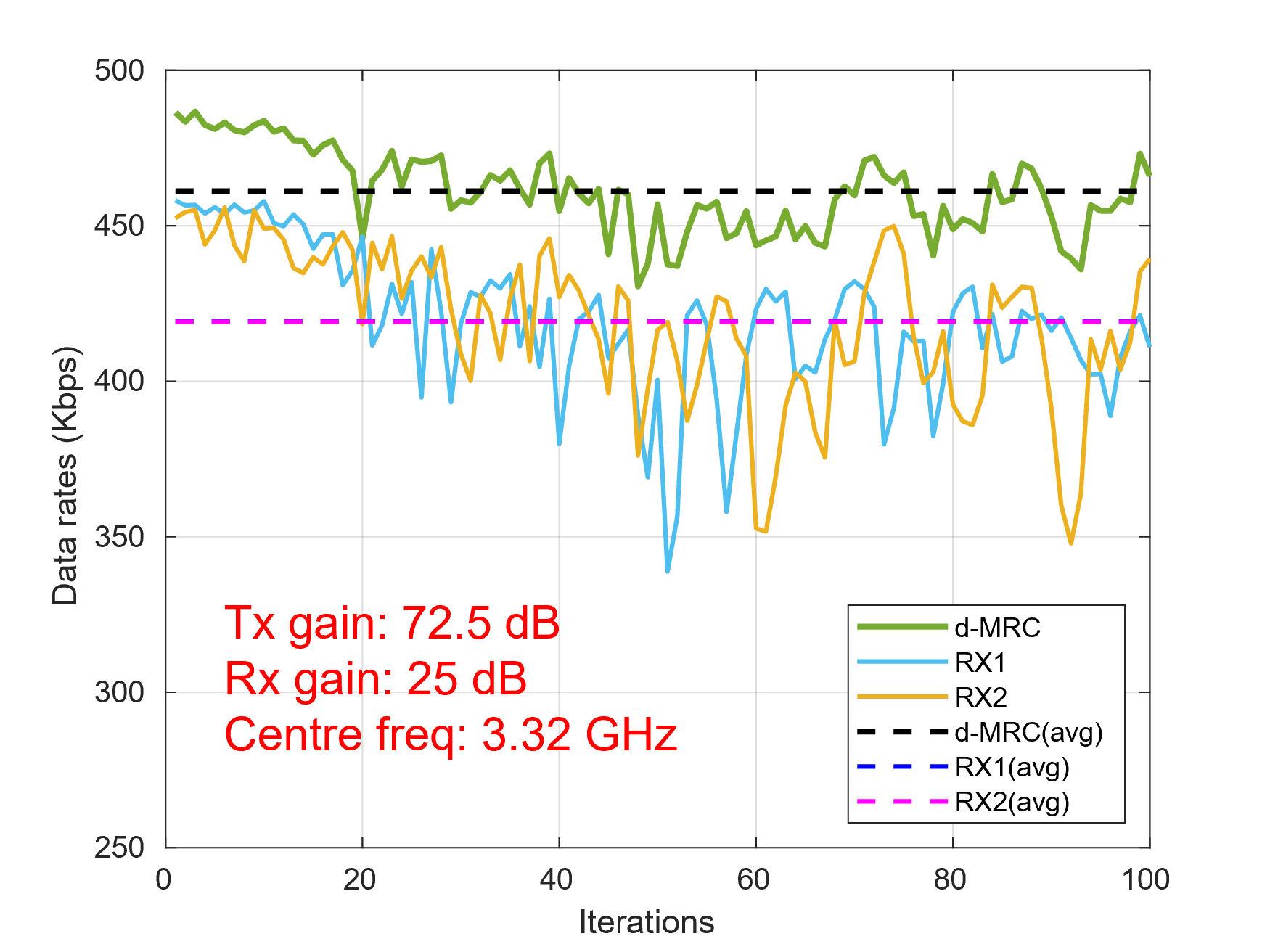}
    \captionsetup{justification=centering}
    \caption{Both UAVs taking path2}
    \label{fig:Datarate - Both UAVs taking path2}
    
    \end{subfigure}
     \captionsetup{justification=centering}
    \caption{Data rates for individual radios and d-MRC}
\end{figure*}

\begin{figure}
\centering
    % \includegraphics[width=0.8\columnwidth]{diagram1.png}
    % \includesvg[inkscapelatex=false,width=0.65\columnwidth]
    \includegraphics[width=0.65\columnwidth]
    {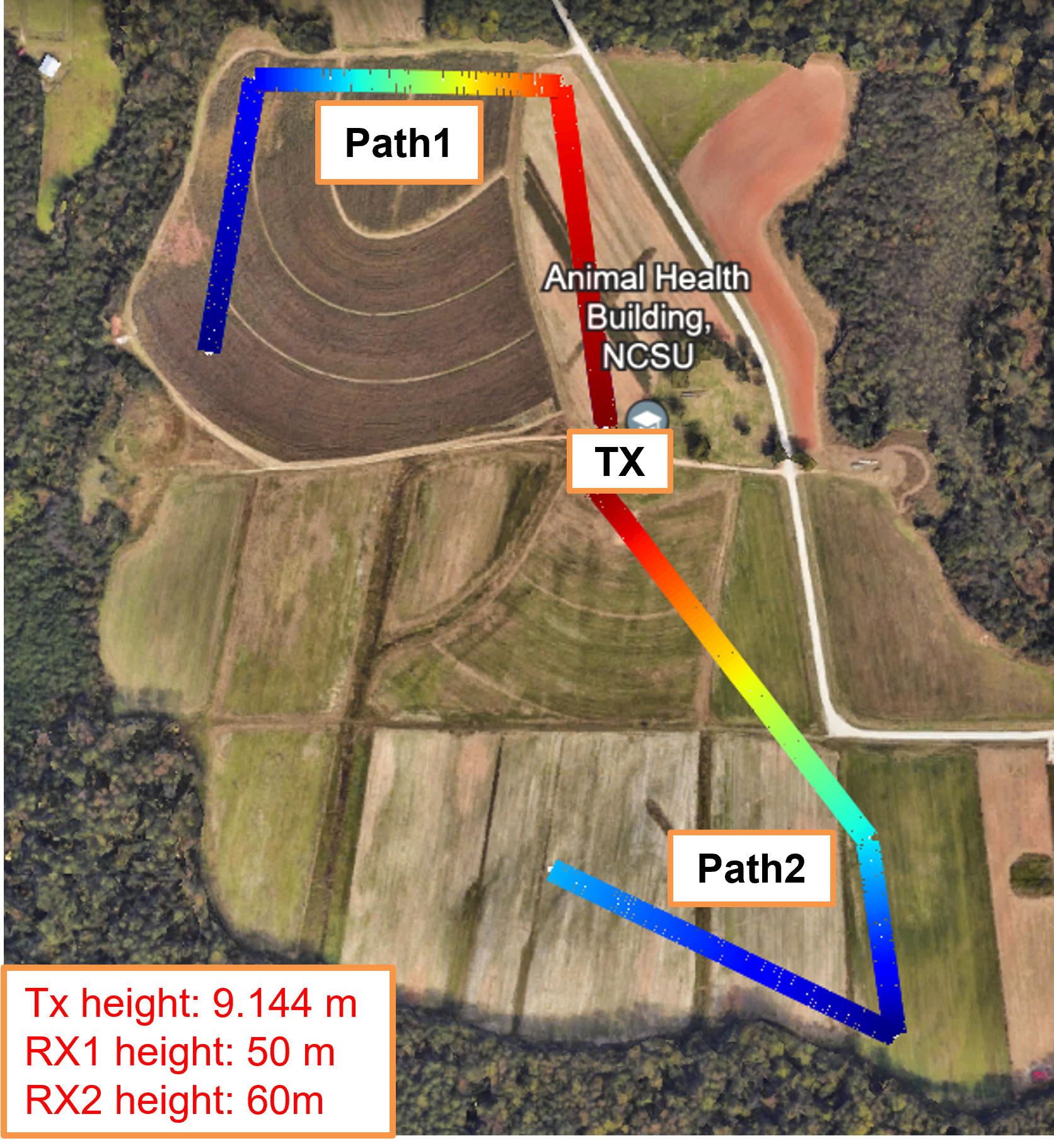}
    \captionsetup{justification=centering}
    \caption{Path for AERPAW experiments.}
    \label{fig:aerpaw_path}
\end{figure}

\subsection{Field Test using AERPAW}
The AERPAW testbed at NC State enables experimental research in a real-world outdoor environment. We leverage this testbed to transmit and receive data, which is then processed offline using our algorithms to ensure that our system is capable of providing good results when implemented in the real outdoor environment. 
% The AERPAW testbed allows us to validate that our system performs well under realistic conditions by exposing our system to practical effects such as doppler shifts and dynamic channel variations. Through these experiments, we demonstrate that our algorithms inherently account for and effectively mitigate these challenges, proving the robustness of our RASHND system in practical scenarios. 
Through these experiments, we demonstrate that our algorithms inherently accounts for and effectively mitigates challenges such as doppler shifts and dynamic channel variations, proving the robustness of our RASHND system in practical scenarios.
The experiment is performed on a $1\times2$ SIMO setup with two moving UAVs acting as receivers and a fixed transmitter. Fig. \ref{fig:aerpaw_path} shows the path taken by the UAVs. We collect three sets of data, first one is for UAV$1$ taking path$1$ and UAV$2$ taking path$2$. For the second set, data is taken when both UAVs take path$1$ and the third set is for both the UAVs moving in path$2$. For all three cases, UAV$1$ is at a height of $50$ m and UAV$2$ is at $60$ m, they move at a speed of $5$ m/s and have a $25$ dB gain. The TX is fixed at a height of $9.144$ m and has a gain of $72.5$ dB. The experiments are conducted with a centre frequency of $3.32$ GHz and a bandwidth of $1$ MHz.

Fig. \ref{fig:Datarate - UAVs taking different paths} - \ref{fig:Datarate - Both UAVs taking path2} shows the individual data rates and d-MRC results for all the three cases described above. It can be seen that the combined results are better than the individual data rates. Each point in x-axis corresponds to the average data rate for $500$ messages and y-axis is the data rates in Kbps. The dotted lines are the average over the entire range.

\section{Conclusion} 
This paper introduces a resilient self-healing network architecture transitioning between the d-MRC, d-LMMSE and SC algorithms to obtain superior data rate values compared to the individual receivers across diverse environments in distributed SIMO systems. We evaluate the performance on bitstream and image data and show that the RASHND self-healing algorithm has superior performance in various environments. RASHND effectively uses D2D communication to share data among the devices demonstrating IDL capabilities. We validate the working of our algorithms in the real-world outdoor environment and also using the AERPAW testbed to emulate UAV performance. We show that the combined data rates are superior than the individual data rates for all the cases.
% We effectively use D2D communication to share data among the devices and prove our ability in having IDL capabilities. We validate the working of our algorithms in the real-world outdoor environment and also using the AERPAW testbed to emulate UAV performance. We show that the combined datarates are superior than the individual datarates for all the cases.

\ifCLASSOPTIONcaptionsoff
  \newpage
\fi

\bibliographystyle{IEEEtran}
\bibliography{references}
\end{document}